# A Bayesian Framework

# For

# Combining Valuation Estimates

## Kenton K. Yee♦

**Mellon Capital Management**
595 Market Street, Suite 3000, San Francisco, CA  94107

**and**

**Columbia Business School**
Uris Hall, 3022 Broadway, New York, NY 10027

kenton@alum.mit.edu

http://papers.ssrn.com/sol3/cf_dev/AbsByAuth.cfm?per_id=240309

**Review of Quantitative Finance and Accounting, Vol. 30.3 (2008)** *forthcoming*

♦ I thank Francis Diebold for a useful discussion about statistical research on combining forecasts, and Jennifer Hersch for editorial assistance.

The statements and opinions expressed in this article are those of the author as of the date of the article and do not necessarily represent the views of Bank of New York Mellon, Mellon Capital Management, or any of their affiliates.  This article does not offer investment advice.

# A Bayesian Framework

# For

# Combining Valuation Estimates

## Abstract

Obtaining more accurate equity value estimates is the starting point for stock selection, value-based indexing in a noisy market, and beating benchmark indices through tactical style rotation. Unfortunately, discounted cash flow, method of comparables, and fundamental analysis typically yield discrepant valuation estimates. Moreover, the valuation estimates typically disagree with market price. Can one form a superior valuation estimate by averaging over the individual estimates, including market price? This article suggests a Bayesian framework for combining two or more estimates into a superior valuation estimate. The framework justifies the common practice of averaging over several estimates to arrive at a final point estimate.

# I. Introduction

Investors, bankers, insiders, analysts, tax authorities, judges, and other capital market participants are constantly pressed to assess value. Even if one believes the market is efficient in the intermediate or long run, there is no reason to believe that instant market quotes reflect long-run value in the short run (Black 1986). On the contrary, market microstructure, liquidity, and behavioral phenomena suggest that prices at any given time may be quite noisy. Indeed, institutional "tactical" asset managers exist to exploit temporary market mispricing and to push the limits of arbitrage. To this end, obtaining more accurate equity value estimates is the starting point for stock selection (Treynor and Black 1973; Penman 2001; Yee 2007), value-based indexing in a noisy market (Chan and Lakonishok 2004; Arnott, Hsu, and Moore 2005; Siegel 2006; Perold 2007), and beating benchmark indices through tactical asset allocation (Black and Litterman 1991 and 1992). Obtaining accurate value estimates is also useful at the corporate finance level. Boards of directors and shareholders determine what prices to tender or accept for mergers and acquisitions. Tax authorities and other regulators must appraise estates, assess the fairness of transfer prices, and audit the value of employee stock option plans (ESOPs). When irreconcilable disputes arise, their assessments are litigated in court by a judge, who chooses a terminal appraisal after weighing conflicting testimony from adversarial experts.

Valuing a company is not rocket science – at least in principle. According to finance theory, value is the sum of expected cash outflows suitably discounted. One way to estimate equity value is the discounted cash flow (DCF) method. DCF is implemented by forecasting expected cash inflows from operations, netting them against forecasted payments to creditors, and then discounting. The discount factor, which depends on interest rates, market risk, and leverage, is estimated from CAPM and the weighted average cost of capital (Kaplan and Ruback 1995). In a certain world with perfect markets, DCF is cut and dry. But owing to uncertainty in prospective cash flows (which may be exacerbated by limited available information) and inherent shortcomings of CAPM (which requires estimating beta and adopting a value for the elusive equity risk premium), DCF analysis inevitably leads to an imprecise answer.

In the face of uncertainty and heterogeneous beliefs or information asymmetry, estimated value – like beauty – is in the eyes of the beholder. A valuation result depends on the procedure one uses to obtain it. Along with DCF, the most popular procedures include asset-based valuation, the method of comparables, and ratio analysis. The method of comparables entails identifying a sample of peer companies and an accounting benchmark such as earnings, revenues, or forward earnings. Assuming the price-to-benchmark ratios of all peer companies are roughly equal, one infers the value of a company from the average or median price-to-benchmark ratio of its peers. Although the method of comparables is arguably the most popular valuation method, it is plagued with subjectivity in both the determination of pertinent comparable measures and the identification of a firm's true peers (Bhojraj and Lee 2003). In practice, the outcome can be manipulated by including or excluding "standouts" or "outliers" from a candidate peer group. Another problem arises when modestly different comparable procedures yield disparate answers when applied to the same company at the same time. No single procedure is conclusively most precise and accurate in all situations.

The financial analysis industry has spawned a dizzying array of competing estimation techniques and variations on each technique. Kaplan and Ruback (1995), Gilson, Hotchkiss, and Ruback (2000), Lie and Lie (2002), Liu, Nissim, and Thomas (2001) and Yoo (2006) ran horse races between different valuation methods. Kaplan and Ruback estimated price-to-EBITDA and also DCF valuations for a sample of highly leveraged transactions (HLTs). For their sample of 51 HLTs between 1983 and 1989, they found price-to-EBITDA and DCF gave comparable levels of accuracy and precision for predicting the HLT transaction price. In particular, about half of their valuation estimates fell within 15% of the actual HLT transaction price. However, many will argue that the glass is half empty. Differences exceeding 15% in the other half of the Kaplan and Ruback sample demonstrates that price-to-EBITDA and DCF are inconsistent about half the time.

Gilson, Hotchkiss, and Ruback (2000) compared DCF to the method of comparables for companies emerging from bankruptcy. Using EBITDA multiples based on the industry median, they found about 21% of the valuation estimates fell within 15% of market values. The authors conjectured that Kaplan and Ruback's HLT valuations were more precise than their bankruptcy ones due to greater information efficiency in the HLT setting. Using 1998 financial data from Compustat, Lie and Lie found that the method of comparables based on asset value yields more precise and less biased estimates than estimates based on sales and earnings multiples, especially for financial firms. Liu, Nissim, and Thomas (2001) and Yoo (2006) found that *forward* price-to-earnings ratios—when consensus analysts' earnings forecasts are available—perform best in large sample studies. Consistent with Liu, Nissim, and Thomas, and Yoo[1] (whose studies are based on a much longer time period), Lie and Lie also conclude that forward price-earnings ratios are more accurate and precise estimators than trailing ones. However, Lie and Lie report that the accuracy and precision of their method of comparable estimates varied greatly according to company size, profitability, and the extent of intangible value.

Since no method universally dominates the others, it is not surprising that financial analysts frequently estimate DCF as well as method of comparable values for a more complete analysis. Accordingly, new valuation textbooks—for example, Copeland, Koller, and Murrin 2000; Damodaran 2002; English 2001; Penman 2001—promulgate multiple techniques and view them as being simultaneously and contextually informative, with no dominating method. When available, market value serves as yet another estimate of value. This article addresses the question, *how should an analyst combine two or more noisy valuation estimates into a superior valuation estimate that makes optimal use of the given information?*

---

[1] Yoo (2006) empirically examined, in a large-scale study of Compustat firms, whether taking a linear combination of several univariate method of comparables estimates would achieve more precise valuation estimates than any comparables estimate alone. He concluded that the forward price-earnings ratio essentially beats the trailing ratios, or combination of trailing ratios, so much that combining it with other benchmark estimates does not help. Yoo did not consider DCF, liquidation value, or market price, in his study.

Although there is tremendous practitioner interest this question, the academic literature offers surprisingly little guidance on how to combine estimates (Yee 2004). When analysts do combine estimates, their techniques are usually ad hoc since there is no theoretical guiding framework. A common practice is to form a linear weighted average of the different valuations.[2]   Indeed, weighted average valuations have become industry practice in financial legal proceedings and IRS tax-related valuation disputes (Beatty, Riffe, and Thompson 1999).   The Delaware Block Method is an important case in point. In the years following the Great Depression, the Delaware Court of Chancery, the leading corporate law jurisdiction in the United States, evolved a valuation procedure that, by the late 1970s, became known as the *Delaware Block Method*. Virtually all American courts employed the Delaware Block Method for appraising equity value in the 1970s and early 1980s. In Delaware, the Block Method was the legally prescribed valuation procedure for shareholder litigation up until 1983. In 1983, the Delaware Supreme Court updated the law to allow judges to consider "modern" methods like discounted cash flow. Subsequently, the Delaware courts currently use the Block Method, discounted cash flow, and hybrids of these methods (Yee 2002; Chen, Yee, and Yoo 2007).

The Delaware Block Method is most easily characterized in algebraic terms, even though judges do not write down such explicit formulas in their legal options. The Block Method relies on three measures to estimate the fair value of equity:

- contemporaneous market price P

- net asset value[3] of equity b

---

[2] The author's equity-valuation students enrolled in an MBA elective at a leading business school like to take weighted averages of their valuation estimates even though nothing in the class materials advises them to combine estimates. When asked why they do this, students report they learned to do this at work, other classes, or that weighing makes common sense. The students do not articulate how they determine their weights, and the weights vary greatly between students.

[3] Net asset value is the liquidation value of assets minus the fair or settlement value of liabilities. Net asset value is not the same as accounting "book" value. Book value, because it derives from historical cost accounting, arcane accounting depreciation formulas, and accounting rules that forbid the recognition of valuable intangible assets, is known to be an unreliable (typically conservatively biased) estimate of net asset value.

- five-year trailing earnings average $\bar{e}$ capitalized by $\varphi$, a number estimated from price-to-earnings ratios of comparable firms.

Then, the fair value **V** of equity at date t is estimated as the weighted average

$$V = (1 - \kappa_b - \kappa_e)P + \kappa_b b + \kappa_e \phi \bar{e}. \qquad (1)$$

The weights $(1 - \kappa_b - \kappa_e)$, $\kappa_b$, and $\kappa_e$ —numbers between 0 and 1 —are adjudicated on a case-by-case basis. Earnings capitalization factor $\phi$ is also determined on a case-by-case basis and typically ranges between 5 and 15.

Even if one accepts combining estimates by averaging them, as in the Delaware Block Method, how should the relative weights on P, b, and $\bar{e}$ be determined? Should more weight be given to market value than net asset value? Should net asset value receive more attention than earnings value? The simplest choice might be to take an equally weighted average, but even this parsimonious rule is ad hoc and unjustified.

On this issue, the IRS valuation rules, honed by decades of spirited high stakes litigation and expensive expert testimony, do not lend insight. Revenue Ruling[4] 59-60 advises that "where market quotations are either lacking or too scarce to be recognized," set $1 - \kappa_b - \kappa_e = 0$ and focus on earnings and dividend-paying capacity. For the comparables method, use the "market price of corporations engaged in the same or similar line of business." Upon recognizing that "determination of the proper capitalization rate presents one of the most difficult problems," the Ruling advises that "depending upon the circumstances in each case, certain factors may carry more weight than others…valuations cannot be made on the basis of a prescribed formula."

Intuition provides a bit more guidance. *The intuitively sensible rule is to weigh the more credible estimates more and the less credible estimates less.* For example, if a security is thinly traded, illiquidity may cause temporary price anomalies.[5]  Likewise, "animal spirits" may cause the price of hotly traded issues to swing unexplainably.  In either extreme, price quotations are unreliable indicators of fundamental value and common sense says to weigh price quotations less in such cases.  On the other hand, price should be given more weight when it is believed that market price is more efficient and has not been subject to animal spirits.

---

[4] The Internal Revenue Code addresses the valuation of closely held securities in Section 2031(b). The standard of value is "fair market value," the price at which willing buyers or sellers with reasonable knowledge of the facts are willing to transact. Revenue Ruling 59-60 (1959-1 C.B. 237) sets forth the IRS's interpretation of IRC Section 2031(b).

This article proposes Bayesian triangulation as the appropriate conceptual framework for combining value estimates.  Section II describes the Bayesian triangulation formula, which prescribes how much weight to give different valuation estimates.   Section III applies Bayesian triangulation to interpret why Delaware Chancellors choose the weights that they do in appraisal hearings.  Section IV discusses the possibility of using regression methods to determine the triangulation weights from historical data, and Section V concludes.

# II. Bayesian Triangulation

This section prescribes how to fold two different valuation estimates together with the market price to achieve a more precise valuation estimate.  First, it introduces Theorem 1 and Eq. (2), the formula prescribing how much to weigh the individual estimates for maximum accuracy of the triangulated valuation estimate.  Then, it describes how to implement Eq. (2) and give a recipe for estimating input parameters.

To state Theorem 1, it is necessary to introduce notation.  Assume the analyst has three distinct valuation estimates: market price P, an intrinsic valuation estimate $V_I$, and a method of comparables estimate $V_C$.   In particular, $V_I$ might be a DCF or net asset valuation estimate, while $V_C$ might be an estimate based on peer-group forward price-to-earnings ratios or trailing price-to-book ratios.  While raw valuation estimates are biased in practice (e.g. Lie and Lie 2002), assume that $V_I$ and $V_C$ are pre-processed to remove anticipated bias. (The theorem does not address the issue of how to remove systematic bias, whose removal is outside the scope of this article.)

---

[5] For example, a large block trade of micro-cap shares may cause temporary price volatility.  Non-synchronous trading may cause apparent excessive price *stability*, as seen in emerging equity markets.

Let **V** denote what the analyst wants to estimate. While it is common to refer to **V** as "intrinsic value" or the "present value of expected dividends," one wants to avoid this trap since "intrinsic value" is not observable. Rather, let us just say that **V** is the value one wants to estimate in a given contextual situation. From an active portfolio management or buy-side financial analysis perspective, **V** represents the price target or "signal" that determines a security's "alpha" parameter in the extended CAPM (Treynor and Black 1973; Grinold 1989). From a tax appraiser's perspective, **V** represents the valuation that is most "fair." From an empirical researcher's perspective (see Section IV), **V** might identified as the future realization of market price at the investment horizon (say, next year's price suitably discounted back to the current date). Then, from this perspective, the question is well defined: "How does one use Bayesian triangulation to make the most accurate forecast of market price (or alpha) at the investment horizon?"

Theorem 1 states how to estimate **V** given inputs $\{P, V_I, V_C, \sigma, \sigma_I, \sigma_C, \rho\}$ where:

- $\sigma$ is the standard error of market price P, which is a noisy measure of fundamental value V. Specifically, $\mathbf{V} = P + e$ where[6] $e \sim N(0, \sigma)$.

- $\sigma_I$ is the standard error of intrinsic valuation estimate $V_I$, also a noisy measure of V. Assume $V_I = \mathbf{V} + e_I$ where $e_I \sim N(0, \sigma_I)$ is uncorrelated with e.

- $\sigma_C$ is the standard error of method of comparables estimate $V_C$, also a noisy measure of V. In particular, $V_C = \mathbf{V} + e_C$ where $e_C \sim N(0, \sigma_C)$ is uncorrelated with $e_I$ but is correlated with market noise so that $\text{corr}(e, e_C) = \rho > 0$.

The amount **e** by which market price P differs from **V** is noise. Since market noise is unobservable, suppose that **e** is an unobservable mean-zero random error. Error **e** reflects

---

[6] $Y \sim N(m, \tau)$ means Y is normally distributed with mean m and standard deviation $\tau$. Note that we leave the possibility that the standard deviations may be heteroskedatic – nothing in our model requires them to be independent of V.

Fischer Black's (1986) assertion that "because there is so much noise in the world, certain things are essentially unobservable."

$V_I$ may represent either a DCF or liquidation valuation estimate of V. Assume its estimation error $e_L$ is uncorrelated[7] with market noise e. In contrast, because method of comparable estimates relies directly on market price, they are affected by temporary market movements. Hence, assume that $e_C$, the method of comparables estimation error, is positively correlated with market noise e. As it is, the key difference between $V_I$ and $V_C$ is that corr($e_1$,e) = 0 while corr($e_C$,e) = ρ > 0.

The central result is:

<u>Theorem 1</u>: *The most precise estimate of value **V** an analyst constructs from a linear combination of P, $V_I$, and $V_C$ is*

$$\bar{V} = (1 - \kappa_I - \kappa_C)P + \kappa_I V_I + \kappa_C V_C, \qquad (2)$$

where

$$k_I = \frac{(1-\rho^2)\sigma^2\sigma_C^2}{(1-\rho^2)\sigma^2\sigma_C^2 + (\sigma^2 + \sigma_C^2 + 2\rho\sigma\sigma_C)\sigma_I^2}$$

$$k_C = \frac{(\sigma + \rho\sigma_C)\sigma\sigma_I^2}{(1-\rho^2)\sigma^2\sigma_C^2 + (\sigma^2 + \sigma_C^2 + 2\rho\sigma\sigma_C)\sigma_I^2}$$

$$(1 - \kappa_I - \kappa_C) = \frac{(\sigma_C + \rho\sigma)\sigma_I^2\sigma_C}{(1-\rho^2)\sigma^2\sigma_C^2 + (\sigma^2 + \sigma_C^2 + 2\rho\sigma\sigma_C)\sigma_I^2}.$$

**<u>Proof:</u>** *See Appendix.*

---

[7] Accounting treatments like mark-to-market accounting or cookie-jar accounting may correlate accounting numbers to market noise. If so, then DCF estimates which rely on accounting numbers to make their cash flow projections may correlate to market noise, so that corr($e_I$,e) ≠ 0. For this reason, the *Appendix* provides the generalization of Theorem 1 to the case when corr($e_I$,e) = $\rho_I$ ≠ 0.

The Appendix also offers a generalization of Theorem 1 to the case when $\text{corr}(e_I, e) = \rho_I \neq 0$, when DCF error $e_I$ correlates to market noise e. In the interest of expository parsimony, assume that $\text{corr}(e_I, e) = 0$ in the main text.

Because the choice of weights $\kappa_I$ and $\kappa_C$ in Theorem 1 minimizes the Bayesian uncertainty of $\hat{V}$, call $\hat{V}$ the *Bayesian triangulation* of P, $V_I$, and $V_C$. In particular, $\hat{V}$ is a more precise estimate than P, $V_I$, or $V_C$ are individually. Bayesian triangulation enables analysts to aggregate the information content of the individual estimates P, $V_I$, and $V_C$ into a more robust estimate, $\hat{V}$, of fundamental value V. Bayesian triangulation combines three different estimates into a more precise valuation estimate, $\hat{V}$.

Eq. (2) has the simple appealing features financial analysts should find reassuring. First, $\hat{V}$ is the most efficient unbiased linear estimator of **V** conditional on P, $V_I$, and $V_C$. In this sense, $\hat{V}$ is the Bayesian rational estimate of fundamental value given P, $V_I$, and $V_C$.

# [INSERT FIGURE 1 NEAR HERE]

If a valuation estimate is exact, it deserves full weight and should be reflected in the expressions of Theorem 1. It is straightforward to verify that if a valuation estimate (P, $V_I$, or $V_C$) is exact, then the expressions in Theorem 1 assign it 100% weight and its counterparts zero weight:

- If $\sigma_I \rightarrow 0$, then $\kappa_I = 1$ and $\kappa_C = (1 - \kappa_I - \kappa_C) = 0$.

- If $\sigma_C \rightarrow 0$, then $\kappa_C = 1$ and $\kappa_I = (1 - \kappa_I - \kappa_C) = 0$.

- If $\sigma \rightarrow 0$, then $(1 - \kappa_I - \kappa_C) = 1$ and $\kappa_I = \kappa_C = 0$.

On the other hand, if a valuation estimate is infinitely imprecise (e.g. worthless), it deserves no weight and the formulas in Theorem 1 assign it zero weight:

- If $\sigma_I \to \infty$, then $\kappa_I = 0$.

- If $\sigma_C \to \infty$, then $\kappa_C = 0$.

- If $\sigma \to \infty$, then $(1 - \kappa_I - \kappa_C) = 0$.

When a valuation estimate is worthless, the two remaining estimates have non-zero weights that depend on their relative precisions.

   If a valuation estimate is noisier, an analyst expects it to receive less weight and the competing estimates to obtain commensurately more weight. As depicted in Figure 1, this is what happens. As the ratio of $\sigma_C$ to $\sigma$ increases, the method of comparables and intrinsic valuation estimates ($\sigma_C = \sigma_I$ in the Figure) become more imprecise relative to market price. Accordingly, when this happens in Figure 1, intrinsic value weight $\kappa_I$ and the method of comparables weight $\kappa_C$ fall while the market value weight $(1-\kappa_I-\kappa_C)$ increases. $\kappa_C$ exceeds $\kappa_I$ in the Figure because the method of comparables estimate, by virtue of being correlated with market price, helps to cancel out market noise and so receives more weight than the uncorrelated intrinsic valuation estimate.

   Recall that $\rho$ is the correlation between market price noise and the method of comparables estimation error, $V_C$. Assuming all firms have positive CAPM beta, the method of comparables error correlates to market noise because, by construction, the former relies on peers' market prices to determine $V_C$. However, the correlation is generally not exact for many reasons, one of which is that the method of comparables is subject to sources of error unrelated to the market price. Such errors may stem from the use of imperfect peers, noisy benchmarks, or questionable benchmarks (e.g. manipulated earnings). Thus, financial analysts cannot expect that $\rho=1$ in realistic situations. However, for the purpose of shedding light on Eq. (2), it is instructive to imagine what would happen in the $\rho=1$ limit. In this limit, algebra reveals that $\lim_{\rho \to 1} \kappa_I = 0$ and Eq. (2) simplifies to $\widehat{V} = \left( \dfrac{\sigma_C}{\sigma + \sigma_C} \right) P + \left( \dfrac{\sigma}{\sigma + \sigma_C} \right) V_C$. In other words, if the only source of error in the method of comparables were market noise, then there is no need to give weight to an independent DCF valuation estimate, $V_I$.  In this idealized world, **V** is perfectly estimated by weighing market price P and method of comparables estimate $V_C$.  Indeed, in this limit the method of comparables error exactly cancels out market noise (perfect diversification).  In real world applications, $\rho \neq 1$, the intrinsic value $V_I$ should always have positive weight, and $\hat{V}$ is an imperfect (though always maximum precision) estimate of V.

In principle, one can apply Eq. (2) by estimating the parameters $\sigma, \sigma_C, \sigma_I$, and $\rho$ based on historical data and then evaluating the weights $\kappa_I$ and $\kappa_C$ based on those estimates.  Alternatively, following Eq. (2) it is possible to run regressions like $1 = (1 - \kappa_I - \kappa_C) + \kappa_I \left( \frac{V_I}{P} \right) + \kappa_C \left( \frac{V_C}{P} \right)$ on appropriate historical data samples for the direct determination of $\kappa_I$ and $\kappa_C$.  Of course, back testing always presumes that the weights are constant over time and whether this is true must be decided on a case-by-case basis.

The main point here is conceptual rather than empirical: *Theorem 1 provides Bayesian justification for the practice of combining individual valuation estimates by taking a linear weighted average.  In such weighted averages, the more reliable estimates command more weight.  Market price does not receive 100% weight unless price is perfectly efficient.*

Finally, Theorem 1 – like valuation generally – should be applied contextually.  As stated, Theorem 1 directly applies only to a continuing firm that is not expected to undergo reorganization or liquidation in the foreseeable future.  If a firm is near bankruptcy and liquidation, then liquidation value is the most relevant estimate of value.  In this context, capitalized earnings, which is a measure of continuing value, does not deserve any weight; even if it might be accurate, it is accurate only as an estimate of continuing value, not of liquidation value.  Similarly, if a firm is expected to be taken over in the near future, then the expected take-over price is the most relevant estimate of value independent of capitalized earnings or net asset value, which are irrelevant in this context.  Context matters.

# III. Triangulation in Forensic Valuation

By improving the precision of valuation estimates, triangulation improves the precision of alpha estimates in the Treynor and Black (1973) approach to active portfolio management. Thus, beating the market is one obvious motive for using triangulation. The triangulation formula may also be used to infer the implied imprecision of valuation estimates from the weights that practitioners assign those estimates. This section focuses on the second application in the context of shareholder appraisal litigation.

Compared to other institutions that appraise equity value, the judicial system is unique since judges promulgate official opinions describing how they arrive at value assessments. Because judicial opinions are supposed to be neutral (unlike the mythically biased sell-side analyst reports), they offer unique first-person accounts of how experienced practitioners perform valuation. More importantly, judicial opinions indirectly reflect the views of the professional financial community because judges solicit testimony from competing appraisal experts before making an appraisal. Judges sometimes appoint a special master or neutral expert to provide an independent, unbiased assessment. Hence, judicial opinions typically benefit from input representing several diverse and sophisticated perspectives. In this light, judicial valuation is an indirect but articulate reflection of how valuation is practiced in the greater financial community.

The Delaware Block valuation method, Eq. (1), is the special case of Eq. (2) when $V_I = b$ and $V_C = \phi \bar{e}$. As described in the Introduction, the courts have not articulated a guiding principle for picking the weights $\kappa_b$ and $\kappa_e$. Indeed, the articulated rule is that "No rule of thumb is applicable to weighting of asset value in an appraisal …; rather, the rule becomes one of entire fairness and sound reasoning in the application of traditional standard and settled Delaware law to the particular facts of each case."[8] As one would expect, judges, after hearing expert testimony, tend to assign more weight to net asset value in liquidation contexts, and more weight to capitalized earnings when valuing a continuing firm. Market price is assigned a weight based on availability and liquidity of the shares, which affect the reliability of price. See Seligman (1984) for vivid case-study descriptions of the facts underlying several sample cases.

---

[8] Bell v. Kirby Lumber Corp., Del. Supr., 413 A.2d 137 (1980)

## [INSERT TABLE 1 NEAR HERE]

If one focuses on valuation outcomes rather than on the details of the legal and valuation process underneath, one readily observes the following trend: net asset value is usually weighted almost twice as much as market value or earnings value. As depicted in Table 1, a survey of a frequently cited sample of judicial appraisals finds that, on average, net asset weight $\kappa_b = 46\%$ --almost twice the average value of the market value and earnings value weights. The weights vary tremendously from case-to-case; the standard deviation of $\kappa_b$ is 26%, more than half of the average value of $\kappa_b$. The same holds for the standard deviations of the market value and earnings value weights.[9]

One explanation for the heavier average weight on net asset value in the tiny Table 1 sample is self-selection. It could be that liquidation or reorganization cases, where break-up or net asset value is more relevant, tend to be adjudicated in court more than going-concern cases. It might also be that cases litigated in court tend to involve companies that do not have liquid shares, or companies that do not have credible earnings numbers. Another explanation is simply that judges, or the experts who testify, trust net asset value more than price or capitalize earnings.

For the sake of illustration, if one pretends that the latter explanation dominates, then it is possible to calculate the implied relative precisions that judges accord the individual value estimates by inverting the formulas in Theorem 1. If judges maximize precision, the average weights in Table 1 imply that, on average, judges view net asset value to be a substantially less noisy estimator of value than even market price. Inverting the relationship between $\kappa_I$ and $\kappa_C$ and $\sigma$, $\sigma_I$, and $\sigma_C$ yields

$$\frac{\sigma_C}{\sigma} = \frac{(1-\kappa_I-2\kappa_C)\rho + \sqrt{(1-\kappa_I-2\kappa_C)^2\rho^2 + 4(1-\kappa_I-\kappa_C)\kappa_C}}{2\kappa_C}$$

and

$$\frac{\sigma_I}{\sigma} = \sqrt{\frac{(1-\rho^2)\kappa_C}{\left(1+\left(\frac{\sigma_C}{\sigma}\right)\rho\right)\kappa_I}} \times \left(\frac{\sigma_C}{\sigma}\right).$$

---

[9] For comparison, a random number homogeneously distributed between 0 and 1 has average value 50% and standard deviation 28.9%. Since the market value and earnings value weights have significantly smaller spread and their means differ from 50%, it does not appear these weights are homogeneously distributed random variables.

Assuming that $\kappa_I \equiv \kappa_b \approx \frac{1}{2}$ and $\kappa_C \equiv \kappa_e \approx \frac{1}{4}$ yields $\sigma_C = \sigma$ and $\sigma_I = \sigma \times \sqrt{\frac{1-\rho}{2}}$. Hence, judges select weights as if they think that (i) market prices are as imprecise as those based on capitalized trailing earnings; and (ii) net asset valuation estimates are more precise than market noise (e.g. $\sigma_I \leq \sigma \times \sqrt{\frac{1}{2}}$).

Again, the purpose of this back-of-the-envelope calculation is to illustrate how one might use historical data together with the framework offered by Theorem 1 to estimate the "implied precisions" of competing value estimates. A definitive calculation of implied precision requires a more comprehensive sample and controls for self-selection and other contextual biases. The next section discusses the possibility of using regression methods for estimating the weight coefficients or, equivalently, the implied precisions.

# IV. Regression Methods

Suppose one uses historical data to evaluate the regression model

$$P_{t+1} = a_P P_t + a_b b_t + a_E E_t + \varepsilon_{t+1},$$

where $P_{t+1}$ is the next-period market price (cum-dividend and suitably discounted), $P_t$ is the current price, $b_t$ is the net asset value estimate, $E_t$ is the capitalized earnings estimate, and $\varepsilon_{t+1}$ is the error. Popular textbooks such as Damodaran (2002, Chapter 18) suggest regressions along similar lines. Then, this regression model is interpretable within the framework of Theorem 1 given $P_{t+1}$ represents the "resolution" of fair value at date $t$. From the perspective Eq. (2), it is expected that $a_P + a_b + a_E = 1$. (A deviation from $a_P + a_b + a_E = 1$ is an indication that one or more of the value estimates $\{P_t, b_t, E_t\}$ is systematically biased over the sample.) Moreover, using the measured values of the regression coefficients, one could invert the weight formulas given in Theorem 1 to deduce the implied relative precisions of $\{P_t, b_t, E_t\}$ for predicting $P_{t+1}$.

The technique of using historical data to estimate parameter values in models like Eq. (2) is, of course, not new. For example, Chan and Lakonishok (2004) and Arnott, Hsu, and Moore (2005) estimate cross-sectional and time-series models to predict future returns based on beginning-of-period weighted averages of book-to-price, earnings-to-price, cash-flow-to-price, and sales-to-price. The literature on using regressions to estimate weights for combining *forecasts* is even more developed. Originating with the seminal paper of Bates and Granger (1969), this literature has sprouted branches that explore behavioral as well as purely statistical aspects of how to optimally combine forecasts. Armstrong (1989), Clemen (1989), and Diebold and Lopez (1996) provide surveys. Borrowing generously from Diebold and Lopez, this section summarizes aspects of this literature that are relevant to combining value estimates.

Combining methods fall roughly into two groups, variance-covariance methods and regression-based methods. Consider first the variance-covariance method, which is due to Bates and Granger (1969). Suppose one has two $s$ steps ahead unbiased forecasts, $\hat{y}_{\tau,s}^{i}$ for $i \in \{1,2\}$. The composite forecast formed as the weighted average

$$\hat{y}_{\tau,s}^{c} = \omega \, \hat{y}_{\tau,s}^{1} + (1-\omega) \hat{y}_{\tau,s}^{2}$$

is unbiased if, and only if, the weights sum to unity (as they do here). The combined forecast error will satisfy the same relation as the combined forecast; that is,

$$\varepsilon_{\tau,s}^{c} = \omega \varepsilon_{\tau,s}^{1} + (1-\omega) \varepsilon_{\tau,s}^{2}$$

with a variance $\sigma_c^2 = \omega^2 \sigma_{11}^2 + (1-\omega)^2 \sigma_{22}^2 + 2\omega(1-\omega)\sigma_{12}$ where $\sigma_{11}^2$ and $\sigma_{22}^2$ are unconditional forecast error variances and $\sigma_{12}^2$ is their covariance. The weight that minimizes the combined forecast error variance is

$$\omega = \frac{\sigma_{22}^2 - \sigma_{12}}{\sigma_{11}^2 + \sigma_{22}^2 - 2\sigma_{12}}.$$

The optimal weight is determined in terms of both the variances and covariances. Moreover, the forecast error variance from the optimal composite is less than $\min(\sigma_{11}^2, \sigma_{22}^2)$.

While this suggests that combining forecasts only improves accuracy, there is a caveat. In practice, one replaces the unknown variances and covariances that underlie the optimal combining weights with consistent estimates. That is, if one is averaging over $L$ in-sample years, one estimates by replacing $\omega$ with $\hat{\sigma}_{ij} = \frac{1}{L} \sum_{\tau=1}^{L} \varepsilon_{\tau,s}^i \varepsilon_{\tau,s}^j$ so that

$$\omega = \frac{\hat{\sigma}_{22}^2 - \hat{\sigma}_{12}}{\hat{\sigma}_{11}^2 + \hat{\sigma}_{22}^2 - 2\hat{\sigma}_{12}}.$$

In the small samples typical for valuation studies, sampling error contaminates the combining weight estimates. Sampling error is exacerbated by co-linearity that normally exists among primary forecasts. Accordingly, that combining weight estimates reduces out-of-sample forecast error is not guaranteed. Fortunately, forecast combination techniques perform well in practice as documented by Clemen's (1989) review.

Next, consider the time-series regression method of forecast combination. In this method, one regresses realizations $y_{\tau,s}$ on previous forecasts of $y_{\tau,s}$ to determine the optimal weights for combining current and future forecasts. In other words, one assumes that

$$\hat{y}_{\tau,s}^c = \omega \hat{y}_{\tau,s}^1 + (1-\omega)\hat{y}_{\tau,s}^2$$

and determines the optimal value of $\omega$ by back testing. Granger and Ramanathan (1984) show that the optimal weight has a regression interpretation as the coefficient vector of a linear projection of realizations $y_{\tau,s}$ onto the forecasts subject to two constraints: the weights sum to unity, and no intercept is included. In practice, one simply runs the regression on available data without imposing these constraints.

The regression method is widely used. There are many variations and extensions. Diebold and Lopez (1996) describe four examples in an excellent review article: time-varying combining weights, dynamic combining regressions, Bayesian shrinkage of combining weights toward equality, and nonlinear combining regressions. The remainder of this section summarizes the examples given by Diebold and Lopez and then warns against the overuse of combining regressions to over-fit in-sample data. For more details and a list of citations to the original literature, refer to the Diebold and Lopez article.

### a. Time-Varying Combining Weights

Granger and Newbold (1973) and Diebold and Pauly (1987) proposed time-varying combining weights, respectively, in the variance-covariance context and in the regression context. In the regression context, one may undertake weighted or rolling combining regressions, or estimate combining regressions with time varying parameters.

Time-varying weights are attractive for a number of reasons. First, different learning speeds may lead to a particular forecast improving over time compared to others. If so, one wants to weight the improving forecast more. Secondly, the design of various forecasting models makes them better forecasting tools in some situations. In particular, a structural model with a well-developed wage-price sector may substantially outperform a simpler model when there is high inflation. In such times, the more sophisticated model should receive higher weight. Third, agents' utilities or other properties may drift over time.

### b. Dynamic Combining Regressions

Serially correlated errors arise in combining regressions. Diebold (1988) considers the covariance stationary case and argues that serial correlation is likely to appear in unrestricted regression-based forecast combining regressions when the unconstrained coefficients do not automatically add up to unity. More generally, serial correlation may capture dynamical elements that are not captured by the explicit independent variables.

Coulson and Robins (1993) point out that a combining regression with serially correlated disturbances is a special case of a combining regression that includes lagged dependent variables and lagged forecasts.

### c. Bayesian Shrinkage of Combining Weights Toward Equality

Fixed arithmetic averages of forecasts are often found to perform well even relative to theoretically optimal composites (Clemen 1989). This is because the imposition of a fixed-weights constraint eliminates variation in the estimated weights at the cost of possibly introducing bias with incorrectly selected weight. However, the evidence indicates that the benefits of imposing equal weights often exceed this cost.

With this motivation, Clemen and Winkler (1986) propose Bayesian shrinkage techniques to allow for the incorporation of varying degrees of prior information in the estimation of combining weights. Least-squares weights and the prior weights correspond to the polar cases of the posterior-mean combining weights. The actual posterior mean combining weights are a matrix-weighted average of the two polar cases. They describe a procedure whereby the weights are encouraged, over time, to drift toward a central tendency (e.g., $\omega = \frac{1}{2}$). While the combining weights are coaxed toward the arithmetic mean, the data is still allowed some opportunity to reflect possibly contrary values.

### d. Nonlinear Combining Regressions

Theory does not specify that combining regressions should be linear. In fact, valuation theory suggests that a degree of nonlinearity is to be expected (Yee 2005). As a practical matter, empiricists favor linear models (only) because of their simplicity, which is always a virtue.

In combining regressions, one way to introduce nonlinearity is to allow the weights to depend on one or more hidden state variables that change dynamically over time. The value of the state variable must be inferred from past forecast errors or the value of other observables. Needless to say, the structural equations can become quite elaborate, the computation quite messy, and the results sensitive to numerical details.

For these reasons, I do not recommend introducing nonlinearity or many variables without specific motivation from economic insight.  Absent sound theoretical guidance, it is too easy for the "exploration" of complex econometric models to turn into fishing expeditions.  Keep in mind that Bayesian regression methods do not offer a license to go on undisciplined fishing expeditions without clear thinking and guidance from economic insight.  Data snooping, even within a Bayesian rubric, is a poor substitute for economic theory if one wants to achieve good out-of-sample performance.

# V. Summary

Bayesian triangulation provides a conceptual framework that justifies two intuitive heuristics:

1. Estimate value as the weighted average of all available *bona fide* valuation estimates, including market price if available.
2. Assign greater weight to the more reliable estimates when some estimates are more reliable than are others.

As presented in Section II, the triangulation formula, Eq. (2), agrees with the Bayesian expectation value conditioned on the information set comprised of the individual valuation estimates.   Applying standard Bayesian analysis, Eq. (2) can be easily extended to provide a conceptual framework for folding in as many valuation estimates as desired.

There are many reasons why obtaining more precise valuation estimates is interesting.  Beating the market with superior portfolio management a la' Treynor and Black (1973) is an obvious motive.  Another reason is more academic.  Section III describes how, given a particular choice of weights, one is able to reverse engineer Eq. (2) to obtain the valuation noise implied by a given set of Bayesian weights.   Reverse engineering reveals that shareholder-litigation judges believe market price is as noisy a valuation estimate as capitalized trailing earnings and significantly noisier than net asset value.   This is interesting because, while conventional wisdom pre-dating Fischer Black has always believed that market prices and valuation estimates are noisy, it is also generally accepted that the amount of noise is an elusive quantity that cannot be observed. Inverting the Bayesian triangulation formula provides a way of estimating the implied noisiness of price and other valuation estimates from a credible sample of appropriate data. This exercise is suggested as an issue for future research.

## Appendix: Proof of Theorem 1

This section establishes that the specific choice of weights $\kappa_I$ and $\kappa_C$ in Theorem 1 maximizes the precision of $\hat{V}$ or, equivalently, minimizes $\text{var}[\hat{V}]$. To this end, consider an arbitrary linear weighing scheme $\hat{V} = \kappa_P P + \kappa_I V_I + \kappa_C V_C$, where $\kappa_P$, $\kappa_I$, and $\kappa_C$ are yet to be determined real numbers. Plugging in $V = P + e$, $V_I = V + e_I$, and $V_C = V + e_C$ and requiring $E[\hat{V}] = V$ immediately yields $\kappa_P = 1 - \kappa_I - \kappa_C$, which implies $\hat{V} = V + (-1 + \kappa_I + \kappa_C)e + \kappa_I e_I + \kappa_C e_C$. Computing the variance of this expression yields $\text{var}[\hat{V}] = (1 - \kappa_I - \kappa_C)^2 \sigma^2 + \kappa_I^2 \sigma_I^2 + \kappa_C^2 \sigma_C^2 - 2(1 - \kappa_I - \kappa_C)\kappa_C \rho \sigma \sigma_C$. The first order conditions for minimizing $\text{var}[\hat{V}]$ with respect to $\kappa_I$ and $\kappa_C$ may be written in matrix form as

$$\begin{pmatrix} \sigma^2 + \sigma_I^2 & (\sigma + \rho\sigma_C)\sigma \\ (\sigma + \rho\sigma_C)\sigma & \sigma^2 + \sigma_C^2 + 2\rho\sigma\sigma_C \end{pmatrix} \begin{pmatrix} \kappa_I \\ \kappa_C \end{pmatrix} = \begin{pmatrix} \sigma^2 \\ (\sigma + \rho\sigma_C)\sigma \end{pmatrix}.$$

Inverting the matrix recovers the expressions for $\kappa_I$ and $\kappa_C$ given in the Theorem. These weights minimize $\text{var}[\hat{V}]$, which means they *maximize* $\left(\text{var}[\hat{V}]\right)^{-1}$, the precision of $\hat{V}$.

**Q.E.D.**

## Generalization of Theorem 1

*Theorem 1 assumes that error $e_I$ from the intrinsic valuation estimate is uncorrelated to market noise e. If an analyst believes that* $\text{corr}(e_I, e) = \rho_I \neq 0$, *he or she should replace the weights in Theorem 1 with the following formulas:*

$$k_I = \frac{\left[(1-\rho^2)\sigma\sigma_C + (\rho\sigma+\sigma_C)\rho_I\sigma_I\right]\sigma\sigma_C}{(1-\rho^2)\sigma^2\sigma_C^2 + 2(\rho\sigma+\sigma_C)\rho_I\sigma\sigma_C\sigma_I + \left[(1-\rho_I^2)\sigma^2 + 2\rho\sigma\sigma_C + \sigma_C^2\right]\sigma_I^2}$$

$$k_C = \frac{\left[(1-\rho_I^2)\sigma\sigma_I + (\rho_I\sigma+\sigma_I)\rho\sigma_C\right]\sigma\sigma_I}{(1-\rho^2)\sigma^2\sigma_C^2 + 2(\rho\sigma+\sigma_C)\rho_I\sigma\sigma_C\sigma_I + \left[(1-\rho_I^2)\sigma^2 + 2\rho\sigma\sigma_C + \sigma_C^2\right]\sigma_I^2}$$

*and*

$$(1-\kappa_I-\kappa_C) = \frac{\left[(\sigma_C+\rho\sigma)\sigma_I + \rho_I\sigma\sigma_C\right]\sigma_I\sigma_C}{(1-\rho^2)\sigma^2\sigma_C^2 + 2(\rho\sigma+\sigma_C)\rho_I\sigma\sigma_C\sigma_I + \left[(1-\rho_I^2)\sigma^2 + 2\rho\sigma\sigma_C + \sigma_C^2\right]\sigma_I^2}$$

**Proof:** Same logic as the proof of Theorem 1. When $\text{corr}(e_I, e) = \rho_I$, the new first order optimization condition becomes

$$\begin{pmatrix} \sigma^2 + \sigma_I^2 + 2\rho_I\sigma\sigma_I & (\sigma + \rho\sigma_C + \rho_I\sigma_I)\sigma \\ (\sigma + \rho\sigma_C + \rho_I\sigma_I)\sigma & \sigma^2 + \sigma_C^2 + 2\rho\sigma\sigma_C \end{pmatrix} \begin{pmatrix} \kappa_I \\ \kappa_C \end{pmatrix} = \begin{pmatrix} (\sigma + \rho_I\sigma_I)\sigma \\ (\sigma + \rho\sigma_C)\sigma \end{pmatrix}.$$

Inverting the matrix recovers the expressions for $\kappa_I$ and $\kappa_C$. These weights maximize the precision of the triangulation estimate $\hat{V}$.

**Q.E.D.**

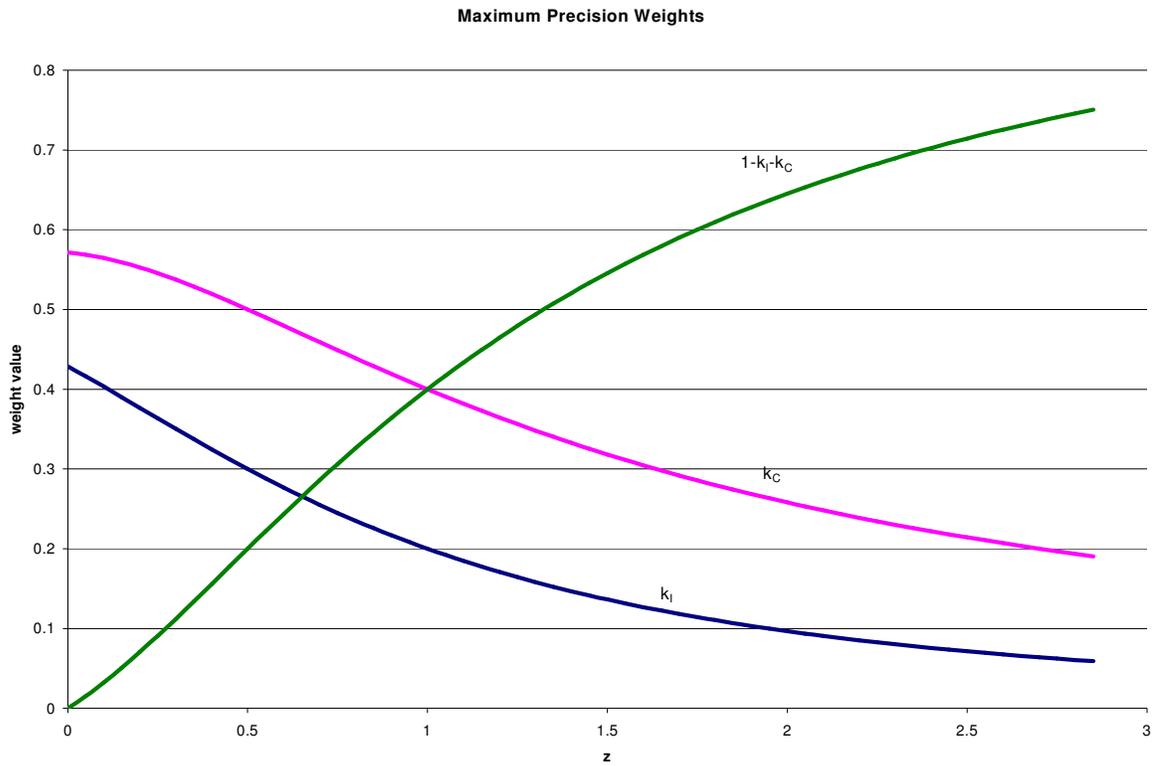

**Figure 1: Triangulation weights $\kappa_C, \kappa_I$, and $1-\kappa_C-\kappa_I$ as a function of $z = \sigma_C/\sigma$ assuming $\rho=0.5$ and $\sigma_C=\sigma_I$.**

| Case | year | $1-k_b-k_e$ | $k_b$ | $k_e$ |
|---|---|---|---|---|
| *Levin v. Midland-Ross* | *1963* | 0.25 | 0.5 | 0.25 |
| *In re Delaware Racing Ass'n* | *1965* | 0.4 | 0.25 | 0.35 |
| *Swanton v. State Guarantee Corp.* | *1965* | 0.1 | 0.6 | 0.3 |
| *In re Olivetti Underwood Corp.* | *1968* | 0.5 | 0.25 | 0.25 |
| *Poole v. N.V. Deli Maatschappij* | *1968* | 0.25 | 0.5 | 0.25 |
| *Brown v. Hedahl's-Q B &R* | *1971* | 0.25 | 0.5 | 0.25 |
| *Tome Land & Improvement Co. v. Silva* | *1972* | 0.4 | 0.6 | 0 |
| *Gibbons v. Schenley Industries* | *1975* | 0.55 | 0 | 0.45 |
| *Santee Oil Co. v. Cox* | *1975* | 0.15 | 0.7 | 0.15 |
| *In re Creole Petroleum Corp.* | *1978* | 0 | 1 | 0 |
| *In re Valuation of Common Stock of Libby* | *1979* | 0.4 | 0.2 | 0.4 |
| *Bell v. Kirby Lumber Corp.* | *1980* | 0 | 0.4 | 0.6 |
| average | | 0.27 | 0.46 | 0.27 |
| standard deviation | | 0.18 | 0.26 | 0.17 |

**Table 1**: These cases are from Seligman (1984). While *Brown*, *Libby*, *Santee Oil*, and *Tome Land* are not Delaware court cases, they all employ the Delaware Block Method.

*End of Document*